\documentclass[iop]{emulateapj} 
\shorttitle{X-ray photoionized bubble in the wind of Vela X-1 pulsar supergiant
companion}
\shortauthors{J.~Krti\v{c}ka, J.~Kub\'at, J.~Skalick\'y}
\usepackage{amsmath,graphicx,amssymb}
\usepackage{natbib}
\usepackage{color}

\begin{document}

\title{X-ray photoionized bubble in the wind of Vela X-1 pulsar supergiant
companion}

\author{Ji\v r\'\i\ Krti\v{c}ka$^1$, Ji\v r\'\i\ Kub\'at$^2$ and
Jan Skalick\'y$^1$}
\affil{$^1$\'Ustav teoretick\'e fyziky a astrofyziky, Masarykova univerzita,
Kotl\'a\v rsk\' a 2, CZ-611\,37 Brno, Czech Republic}
\affil{$^2$ Astromomick\'y \'ustav Akademie v\v{e}d \v{C}esk\'e republiky,
Fri\v{c}ova 298, CZ-251 65 Ond\v{r}ejov, Czech Republic}

\begin{abstract}
Vela X-1 is the archetype of high-mass X-ray binaries, composed of a neutron
star and a massive B supergiant. The supergiant is a source of a strong
radiatively-driven stellar wind. The neutron star sweeps up this wind, and
creates a huge amount of X-rays as a result of energy release during the process
of wind accretion. Here we provide detailed NLTE models of the Vela X-1
envelope. We study how the X-rays photoionize the wind and destroy the ions
responsible for the wind acceleration. The resulting decrease of the radiative
force explains the observed reduction of the wind terminal velocity in a
direction to the neutron star. The X-rays create a distinct photoionized region
around the neutron star filled with a stagnating flow. The existence of such
photoionized bubbles is a general property of high-mass X-ray binaries. We
unveiled a new principle governing these complex objects, according to which
there is an upper limit to the X-ray luminosity the compact star can have
without suspending the wind due to inefficient line driving.
\end{abstract}


\section{Introduction}

A high-mass X-ray binary (HMXB) is a binary star system consisting of a massive
luminous hot star (frequently OB supergiant) and a compact object, either a
neutron star or a black hole \citep{rem}. A fraction of the stellar wind of the
luminous hot star is trapped in the gravitational well of the compact object,
and is accreted onto its surface \citep{davos,laheupet}. Part of the released
potential energy of accreting material is transformed into X-rays, resulting in
one of the most powerful stellar X-ray sources. Such systems of stars in
interaction belong to the most valuable astrophysical laboratories. The binary
nature of the object enables us to determine stellar parameters precisely, which
subsequently serve as a firm base for further study.

Vela X-1 (HD\,77581, GP\,Vel) is the archetype of high-mass X-ray binaries,
consisting of a neutron star and a massive B supergiant
\citep{chodil,brucato,barziv,tom}. The neutron star is a source of pulsed X-ray
and $\gamma$-ray emission with a period of 283\,s \citep{mekac,sever}, modulated
both by the orbital motion and stochastic variations \citep{bildik}. The X-rays
propagating through the hot star wind probe the wind structure, yielding
information about the mass-loss rate and the velocity field \citep{viteal}. The
perpetual X-ray variation (flaring) reveals the existence of some structure in
the wind -- clumping \citep{ducim,prvni}. On the other hand, X-rays also
significantly influence the stellar wind, resulting in X-ray photoionization of
its material \citep{mav,viteal}. Because the stellar wind of hot stars is mostly
driven by the light absorption in the lines of heavier elements, the X-ray
photoionization may influence the wind acceleration. Particularly, the
appearance of highly charged ions, which absorb the light less effectively than
low-charged ions, causes the decrease of the radiative force. Since this force
is responsible for driving the wind, the wind flow may subsequently stagnate.

Numerical studies of stellar winds in HMXBs concentrate mainly on the
multidimensional simulation of wind accretion \citep{blok,blowoo,felan,hadvitr},
while
the wind driving is simplified using force multipliers that take the X-ray
irradiation into an account in an approximative way \citep{stekal,stesam}. This
is a significant shortcoming, because the X-ray photoionization affects the
radiative force, and consequently the amount and velocity of wind material
accreted on the compact companion. Detailed modelling of ionization and
excitation balance in the wind is crucial for the understanding of the influence
of X-ray photoionization on the wind dynamics. The ionization and excitation
balance should be properly derived using equations of statistical equilibrium.%
\footnote{%
This approach is usually referred to as non-LTE or NLTE and it means that the
assumption of thermodynamic equilibrium is not used for evaluation of the
excitation and ionization balance.}
To remedy the situation, we provide wind
models that include the influence of X-ray irradiation using up-to-date NLTE
models.

\section{Vela X-1 primary wind model}

The applied models of the Vela X-1 primary wind are based on the NLTE code
with comoving frame (CMF) line force \citep{cmf}. Our models enable to
selfconsistently predict wind structure just from the stellar parameters (the
effective temperature, mass, radius, and chemical composition). Here we assume
that the stellar wind of the supergiant component is symmetric with respect to
the binary axis (connecting centers of both components) and that the stellar
wind in the direction given by the inclination $\phi$ from the binary axis can
be locally described by a spherically symmetric wind model. The influence of
the neutron star is taken into account by its inclusion as the source of
external X-ray irradiation of the wind.

\begin{table*}[t]
\caption{Parameters of Vela X-1 (HD\,77581) binary system}
\label{hvezpar}
\begin{center}
\begin{tabular}{ccc}
\hline
\hline
Parameter & Value & Reference \\
\hline
\multicolumn{3}{c}{Binary}\\
\hline
Separation $D$& $53.4\,\mathrm{R}_\odot$ & \cite{vani}\\
Period $P$& $8.96\,\mathrm{day}$ & \cite{vani}\\
\hline
\multicolumn{3}{c}{Supergiant}\\
\hline
Spectral type & B0.5Iae & \\
Radius $R_*$ & $30\,\mathrm{R}_\odot$ & \cite{vani}\\
Mass $M$ & $23.5\,\mathrm{M}_\odot$ & \cite{vani} \\
Effective temperature $T_\mathrm{eff}$ & $27\,000\,\mathrm{K}$ & \cite{straku}\\
Wind mass-loss rate $\dot M$ & $1.5\times10^{-6}\,\mathrm{M}_{\odot}\,\mathrm{year}^{-1}$ & this work\\
Wind terminal velocity $v_\infty$ & $750\,\mathrm{km}\,\mathrm{s}^{-1}$ & this work\\
\hline
\multicolumn{3}{c}{Neutron star}\\
\hline
Mass & $1.88\,\mathrm{M}_\odot$ & \cite{vani} \\
X-ray luminosity $L_\mathrm{X}$ & $3.5\times10^{36}\,\mathrm{erg}\,\mathrm{s}^{-1}$ &
\cite{viteal}\\
\hline
\end{tabular}
\end{center}
\end{table*}

Basic parameters of Vela X-1 binary system are given in Table~\ref{hvezpar}.
Binary parameters and the physical parameters of binary members are taken from
the spectroscopic analysis \citep{vani}. The effective temperature of the
supergiant is taken from the tables of \cite{straku} for the corresponding
spectral type. The derived value relatively well agrees with the determination
based on NLTE model atmospheres \citep{fraser}. The parameters of the X-ray
source are adopted from observational analysis of \cite{viteal}. For our
calculations we assume the solar chemical composition \citep{asp09}.

\subsection{Wind model without X-ray irradiation}
\label{bezx}

For supergiant parameters given in Table~\ref{hvezpar} we first calculated NLTE
wind model with CMF line force neglecting the influence of the companion star.
The emergent surface flux is taken from the H-He spherically symmetric NLTE
model stellar atmospheres \citep{kub}. The predicted wind mass-loss rate
$1.5\times10^{-6}\,\mathrm{M}_{\odot}\,\mathrm{year}^{-1}$ well agrees with its
estimate derived from the observed X-ray spectrum
$1.5-2\times10^{-6}\,\mathrm{M}_{\odot}\,\mathrm{year}^{-1}$ \citep{viteal}. The
predicted terminal velocity $750\,\mathrm{km}\,\mathrm{s}^{-1}$ is lower than
the observed one $1100\,\mathrm{km}\,\mathrm{s}^{-1}$ \citep{prina}. This
disagreement may stem either from the model simplifications, inaccurate stellar
parameters (e.g., metallicity), or from the sensitivity of the wind terminal
velocity to a detailed ionization balance in the outer wind regions
\citep{pusle}, which probably manifests itself in a significant scatter of the
ratio of the terminal to the escape velocity for the stars with the same
effective temperatures \citep[e.g.,][]{zelib}.

The velocity structure of the wind model without X-ray irradiation can be
approximated by
\begin{multline}
\label{vrfit}
\tilde v(r)=\left[{v_1\left({1-\frac{R_*}{r}}\right)
                  +v_3\left({1-\frac{R_*}{r}}\right)^3}\right]\times\\
   \left\{{1-\exp\left[{\gamma\left({\frac{r}{R_*}-1}\right)^2}\right]}\right\},
\end{multline}
where
\begin{equation}
v_1=1042\,\mathrm{km}\,\mathrm{s}^{-1},\quad
v_3=-297\,\mathrm{km}\,\mathrm{s}^{-1},\quad
\gamma=-1220.
\end{equation}
Note that the representation  of the velocity law by a polynomial expansion
provides better approximation than the ordinary ``$\beta$-velocity law'',
because these polynomials may form a functional basis \citep{beta}. The
exponential term is included for a better fit of the velocity law close to the
sonic point. The calculated X-ray opacity per unit mass averaged for radii
$1.5\,R_*-5\,R_*$ may be approximated by
\begin{equation}
\label{kapafit}
\tilde \kappa^\mathrm{X}(\nu)=
  \left\{\begin{array}{c}
    a_1(\lambda-b_1)^2,\quad \lambda<\lambda_1,\\
    a_2(\lambda+b_2)^3,\quad \lambda>\lambda_1,\\
  \end{array}\right.\\
\end{equation}
where $\lambda=10^8c/\nu$, $a_1=0.704\,\mathrm{g}^{-1}\,\mathrm{cm}^2$,
$b_1=1.056$, $a_2=4.06\times10^{-3}\,\mathrm{g}^{-1}\,\mathrm{cm}^2$,
$b_2=11.41$, and $\lambda_1=20.18$. We stress that $\lambda$ enters as a
non-dimensional parameter here, which has for a convenience the same value as
the wavelength in units of \AA.

\subsection{Modelling of the 2D wind structure}

The full treatment of the problem would essentially require a complex solution
of 3D time-dependent hydrodynamic equations \citep{frc,blok,felan}. However,
because the typical wind flow-time $R_*/v_\infty\approx0.3\,\mathrm{day}$ is
roughly by a factor of 30 shorter than the orbital period, we neglect the
influence of the orbital motion on the wind structure. Consequently, we assume
that the stellar wind of the supergiant component is axisymmetric with respect
to the binary axis connecting centers of components. Moreover, we assume that
the stellar wind in the direction given by the inclination $\phi$ from the
binary axis can be modelled by a spherically symmetric wind (see
Fig.~\ref{taut}). Consequently, in the orbital plane of the binary our 2D wind
model consists of sectors of a circle.

The wind model in each sector is calculated using our NLTE wind code. Because
the radial wind velocity may be non-monotonic in some sectors, we do not use the
CMF line force for the calculation of these models directly, however the line
radiative force is given by the force calculated in the Sobolev approximation
\citep[e.g.,][]{cassob} multiplied by the ratio of the CMF to the Sobolev line
force derived from the wind model that neglects the radiation from the companion
star (Sect.~\ref{bezx}).

The influence of the neutron star is taken into account only by inclusion of its
X-ray radiation due to the wind accretion on the neutron star. This radiation
irradiates the supergiant and interacts with its wind. To describe this effect,
we add an additional term $J^\mathrm{X}(\nu,r,d)$ to the specific intensity
$J(\nu,r,d)$ in the form
\begin{equation}
\label{xneutron}
J^\mathrm{X}(\nu,r,d)=\frac{L^\mathrm{X}(\nu)}{16\pi^2d^2}\mathrm{e}^{-\tau(\nu,r,d)},
\end{equation}
where the optical depth along the given ray is
($z$ measures the distance along this ray)
\begin{equation}
\label{zamlhou}
\tau(\nu,r,d)=\int_0^d\kappa(\nu,z)\rho(z)\mathrm{d} z,
\end{equation}
$L^\mathrm{X}(\nu)$ is the X-ray luminosity per unit of frequency, $d$ is the
distance from the given point in the supergiant wind region to the surface of
the neutron star, $\kappa(\nu,z)$ is the absorption coefficient per unit of
mass, and $\rho(z)$ is the wind density. The distribution of emergent X-rays
$L^\mathrm{X}(\nu)$ is approximated by the power law \citep{viteal}. Energies
higher than 3~keV, which are well above the ionization energies of all included
ions were not considered in the model. The absorption coefficient and the
density in Eq.~\eqref{zamlhou} can be derived from models for individual
sectors. However, to simplify our approach, for the calculation of
$J^\mathrm{X}(\nu)$ we use fits following from Eqs.~\eqref{vrfit},
\eqref{kapafit}
\begin{align}
\rho(z)=&\frac{\dot M}{4\pi r^2 v(z)},\\
v(z)=&\min(\tilde v(r),\hat v(\phi)),\\
\kappa(\nu,z)=&\tilde \kappa^\mathrm{X}(\nu),
\end{align}
where the relation between the distance along the ray $z$ and the radius $r$ is
derived from the geometry of the problem, and $\hat v(\phi)$ is an average
velocity in the velocity plateau which occurs due to X-rays (see
Fig.~\ref{rychl}).

The models describing the wind in different sectors with inclination $\phi$ with
respect to the binary axis are calculated for a sequence in $\phi$ with a
step of $10^\circ$.

\section{Wind structure in 2D}

\begin{figure}[t]
\centering
\resizebox{0.9\hsize}{!}{\includegraphics{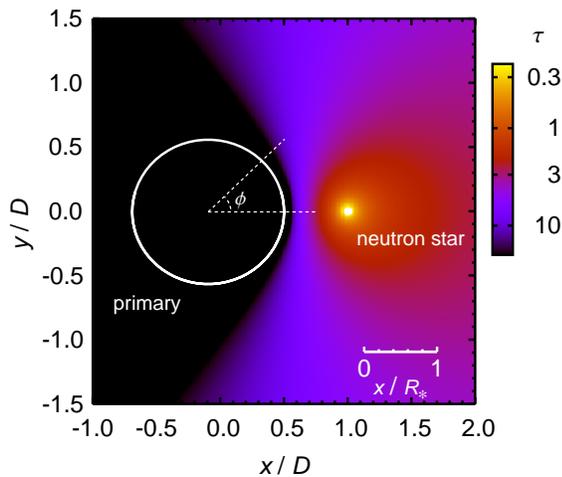}}
\caption{The optical depth for $830\,\mathrm{eV}$ between the neutron star and a
given point plotted in the plane containing the binary axis. The distances are
given in the units of binary separation $D$ with a small ruler plotted in units
of supergiant radius $R_*$. Note that the optical depth plotted in the figure is
only an approximative one, because it was calculated using fits
Eqs.~\eqref{vrfit}, \eqref{kapafit} to wind solution without any external X-ray
irradiation. Anyway, the main trends are the same as in a more precise 2D model.
}
\label{taut}
\end{figure}

The X-ray source located on the surface of the neutron star influences the
ionization state of the supergiant wind. The influence is stronger if the X-ray
optical depth between a given point in the wind and the neutron star surface is
lower. To illustrate this, in Fig.~\ref{taut} we plot the optical depth
(Eq.~\eqref{zamlhou}) in a plane containing the binary axis.
Due to the assumed
symmetry of the problem, the optical depth is axisymmetric with respect to the
binary axis. The X-rays strongly penetrate the wind that directly faces the
neutron star. On the other hand, due to geometrical reasons, the radial wind
streams that are significantly inclined with respect to the neutron star are
affected by X-ray radiation at larger distances from the primary.

\begin{figure}[t]
\centering
\resizebox{0.9\hsize}{!}{\includegraphics{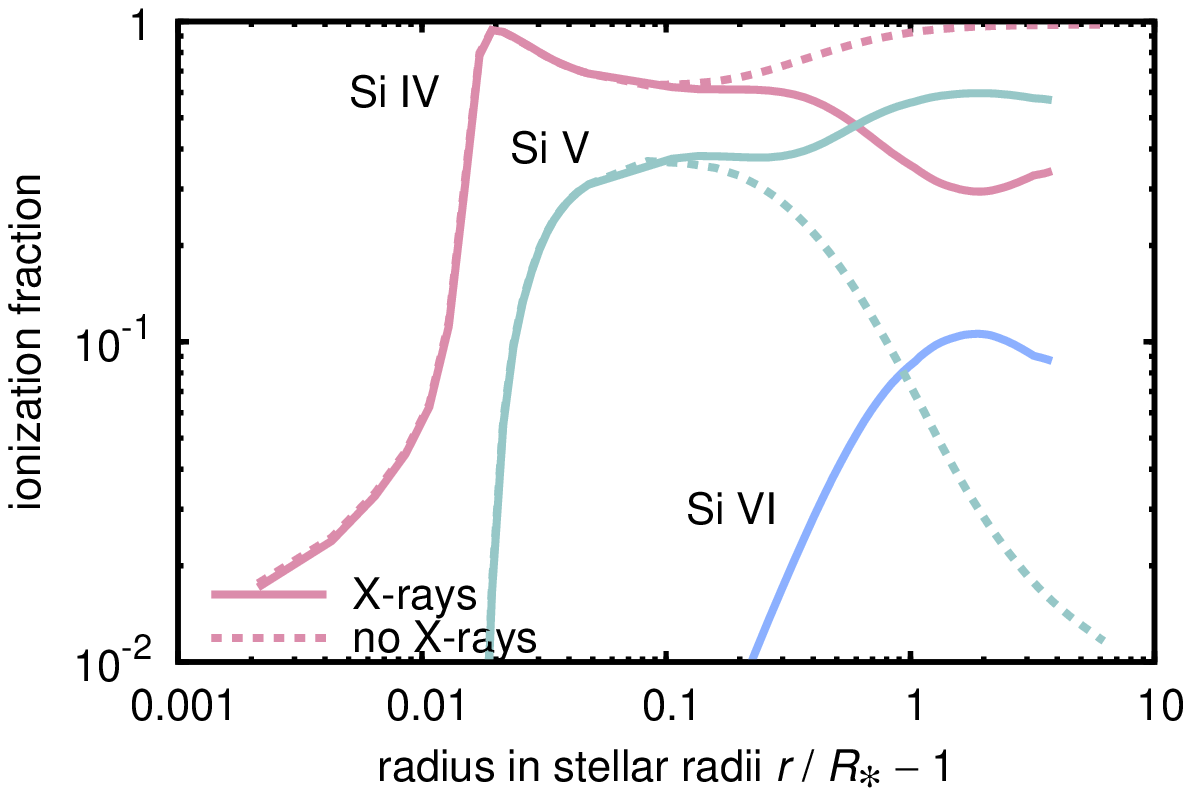}}
\resizebox{0.9\hsize}{!}{\includegraphics{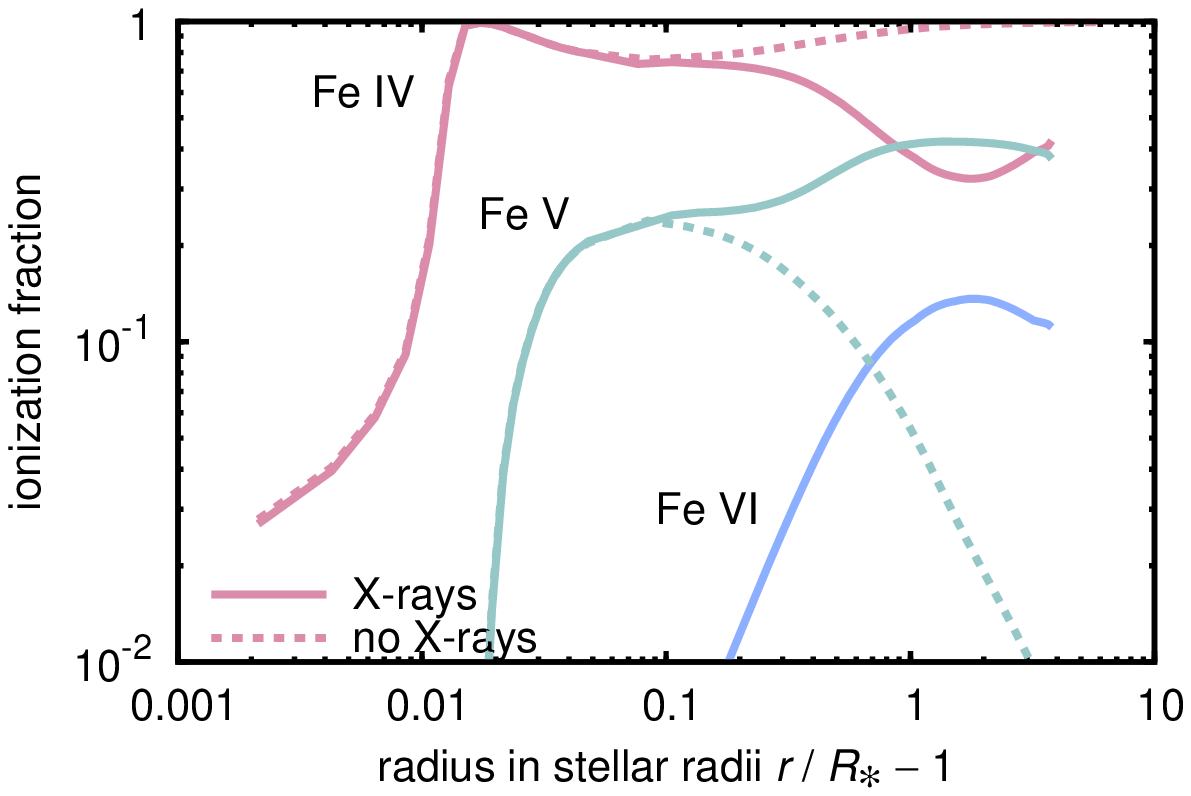}}
\caption{Influence of X-ray photoionization on the ionization state of selected
elements. Calculated for a sector with $\phi=50^\circ$.}
\label{ion}
\end{figure}

As a result of the X-ray photoionization of the wind, lower ionic states are
effectively destroyed and higher ionic states appear in a nonnegligible amount.
This can be seen from Fig.~\ref{ion}, where we compare the ionization fraction
of selected ions in the model with and without X-ray irradiation. The X-rays
influence the wind ionization state in the region where the X-ray optical depth
between a given point and the neutron star is low, $\tau\lesssim1$, i.e, to a
distance comparable to an orbital separation $D$. X-rays are not able to
penetrate the wind close to the supergiant star (since $\tau\gg 1$ if we aim to
approach the supergiant surface), consequently the ionization state of material
there is not affected by X-rays.

\begin{figure}[t]
\centering
\resizebox{0.9\hsize}{!}{\includegraphics{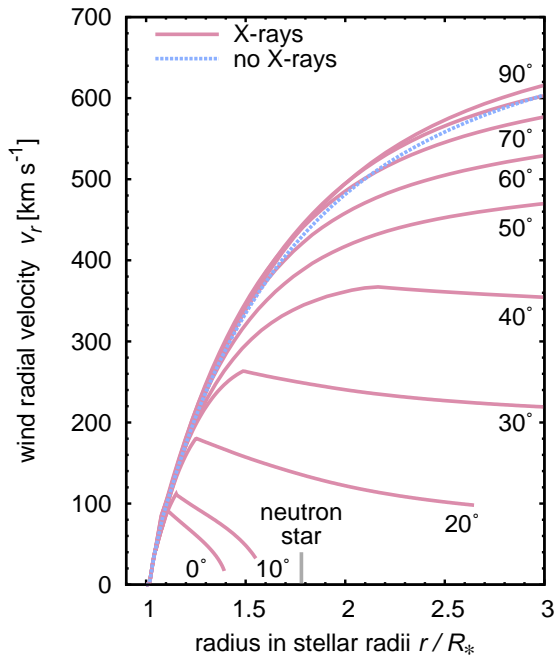}}
\caption{The radial wind velocity for different inclinations $\phi$ with respect
to a neutron star. The position of the neutron star is marked in the graph.}
\label{rychl}
\end{figure}

Because the stellar wind of hot stars is accelerated by the light absorption in
the lines of heavier element ions, any change of the wind ionization state
affects the accelerating radiative force. On average, ions with higher charge
are less effective in driving the wind, basically due to the lower number of
their spectral lines. For a weak X-ray irradiation the emergence of new
ionization states causes a slight increase of the radiative force
\citep[cf.][]{nlteiii}. On the other hand, for a strong irradiation when the
degree of ionization is higher the radiative force significantly decreases
\citep{mav}. Because this effect appears in the outer wind region and the wind
mass-loss rate is determined in the inner wind region, only the wind velocity
(see Fig.~\ref{rychl}), and not the wind mass-loss rate \citep{mav} is affected
by X-ray photoionization. This is especially apparent for lower wind
inclinations ($\phi<40^\circ$), where the wind radial velocity becomes
nonmonotonic. The decrease in the radiative force is so strong there that the
accelerating wind solution is not possible any more, and the wind velocity
switches to decelerating overloaded solutions \citep{feslop} with a typical kink
in the velocity profile. This has further consequences for even lower wind
inclinations. The radiative force in the rays that closely pass the neutron star
($\phi<15^\circ$) is so strongly affected by the X-ray photoionization that the
radiatively driven wind in these directions never reaches the neutron star.

As the result of X-ray photoionization, a region (bubble) around the neutron
star is created, in which the wind stagnates at low velocities, and consequently
high densities. These results explain the detection of low-speed wind regions in
the direction of the neutron star by \cite{viteal}.

\section{Discussion}

\subsection{X-ray luminosity}

Part of the mass which leaves the primary star via its wind can be accreted onto
the compact component. Its potential energy is then released in the form of
X-ray radiation. Let us adopt an approximative Hoyle-Lyttleton treatment to
estimate the X-ray luminosity. For moving medium, the radius from which the
matter can be accreted onto a point mass is given by the relation
$r_\mathrm{HL}={2GM_\mathrm{X}}/{v^2}$, where $M_\mathrm{X}$ is the mass of the
compact component (neutron star), $v^2=v_\mathrm{wind}^2+v_\mathrm{orb}^2$,
$v_\mathrm{orb}$ is the orbital velocity of the compact component, and
$v_\mathrm{wind}$ is the wind velocity at the distance of the compact component
(denoted by $D$). Given $v_\mathrm{orb}=300\,\mathrm{km}\,\mathrm{s}^{-1}$ and
our simulated wind structure, we find that the neutron star is able to collect
matter only from a narrow cone defined by the value of $\phi< 15\,^\circ$ (for
$\phi=15\,^\circ$ and the distance $D$ we have
$v_\mathrm{wind}=32{.}0\,\mathrm{km}\,\mathrm{s}^{-1}$ and
$r_\mathrm{HL}=0{.}17D$). We can express the relation between the accretion rate
$\dot{M}_\mathrm{acc}$ and the mass loss rate $\dot{M}$ from the supergiant as
\citep{viteal} $\dot{M}_\mathrm{acc}=\dot{M} {r_\mathrm{HL}^2}/(4D^2)$. Then the
X-ray luminosity $
L_\mathrm{X}={GM_\mathrm{X}\dot{M}_\mathrm{acc}}/{R_\mathrm{X}} $ is
$L_\mathrm{x}=8{.}8\times 10^{37}\,\mathrm{erg}\,\mathrm{s}^{-1}$. This value is
approximately one order of magnitude higher than the observed one
\citep{viteal}. The reason of this difference is the fact that only a small
fraction of matter in the accretion cone gets finally accreted and the rest
falls back to the surface of the primary as a consequence of the flow stagnation
between the supergiant and the neutron star.

\subsection{Existence of two types of solutions}
\label{nasebista}

The wind equations allow the existence of two types of solutions giving
different X-ray luminosities and wind velocities. The solution presented here
appears in the case of strong X-ray source, which significantly affects the wind
ionization state and consequently also the radiative force. This results in a
slow wind that can be accreted by the neutron star in large amounts producing a
strong X-ray source.

Another type of solution may occur in the case of a weak X-ray source that does
not significantly influence the wind ionization state. The radiation force
becomes higher, similar to that without any X-ray irradiation. This results in
faster outflow roughly corresponding to the ``no X-rays'' case in
Fig.~\ref{rychl}. Due to the dependence of the accretion rate on the velocity
via $\dot{M}_\mathrm{acc}\sim v^{-4}$ and having roughly two times higher $v$,
the X-ray luminosity is by an order of magnitude lower in this case (since
$L_\mathrm{X}\sim \dot{M}_\mathrm{acc}$).
This agrees with adopted assumption of weak X-ray source.
Consequently, also this is a possible
solution of wind equations.

Different types of solutions may appear in different systems, or even an
external perturbation may cause switching between these two wind solutions in a
particular binary system, possibly contributing to the variability of X-ray
luminosity. This variability might be accompanied by the variability of the
distribution of emitted X-rays, if the accretion regime changes with accretion
rate \citep{nerozumim}.

\subsection{Implication for other HMXBs}

\begin{figure}
\centering
\resizebox{0.9\hsize}{!}{\includegraphics{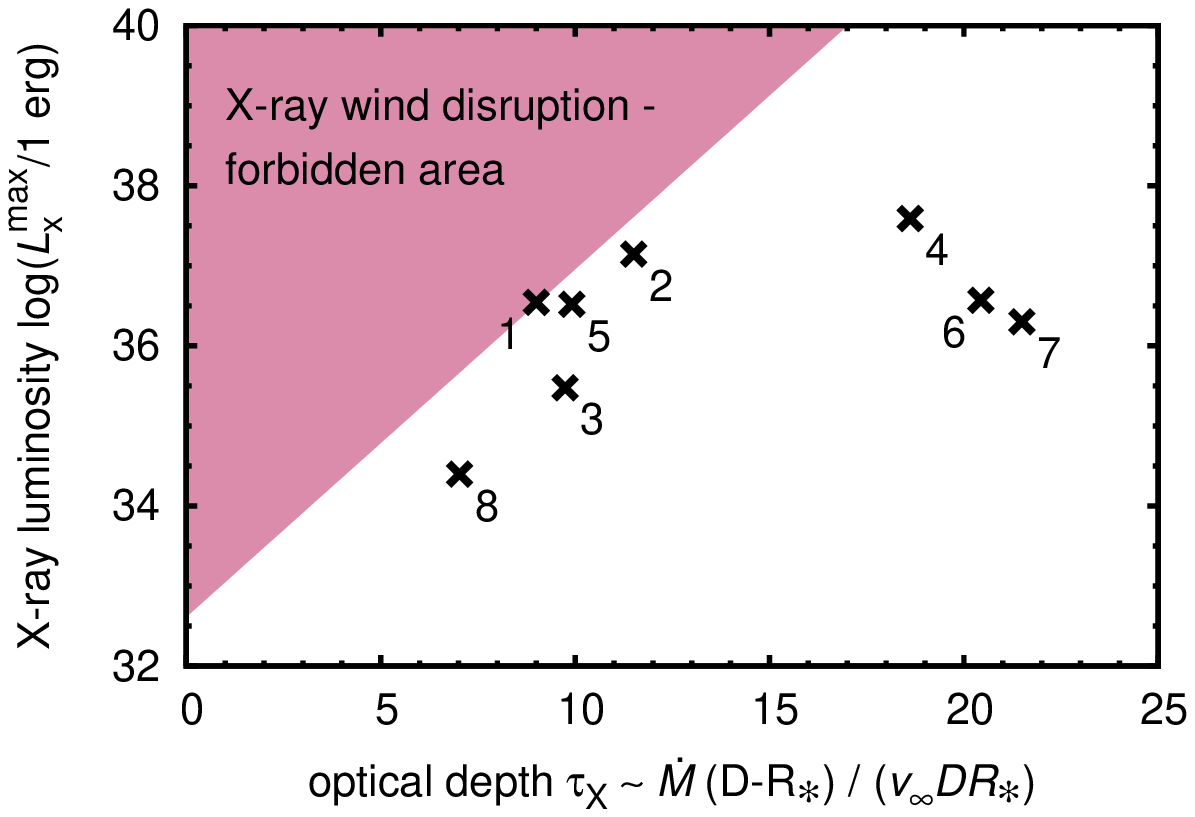}}
\caption{The location of forbidden X-ray luminosities as a function of the
optical depth parameter 
(Eqs.~\eqref{hloubkane}, \eqref{zakladnismernice}).
Observed parameters of
individual HMXBs (crosses) are located outside the highlighted forbidden area.
Here we plotted points (denoted by numbers) for binaries Vela X-1
\citep[1,][]{viteal}, Cyg X-1 \citep[2,][]{ylhad}, IGR J19140+0951
\citep[3,][]{prat}, Wray 977 \citep[4,][]{kaper},  4U 2206+54
\citep[5,][]{rybka}, 4U 0114+65 \citep[6,][]{reig,muk}, 4U 1907+09
\citep[7,][]{cokam}, and LS 5039 \citep [8,][] {cariri}. The plot of the maximum
luminosity was derived using the optical depth between the neutron star and the
wind critical point $\tau_\mathrm{X}(\mbox{Vela X-1})=9$.}
\label{hmxb}
\end{figure}

For a slightly lower mass-loss rate or for a slightly higher X-ray luminosity
than assumed here the X-rays could penetrate deeply into the stellar wind and
significantly influence the ionization state at the wind base resulting in the
decrease of the radiative force. This could lead to the disruption of the
stellar wind and a significant decrease of X-ray luminosity, possibly providing
another contribution to the overall X-ray variability observed in Vela X-1
\citep{krey}.

This means that there exists a maximum X-ray luminosity the compact star can
have for a given geometry of the system and wind mass-loss rate. Assuming that
the Vela X-1 luminosity is the maximum one, the maximum allowed X-ray
luminosities for other HMXBs $L_\mathrm{X}^\mathrm{max}$ can be derived using
the optical depth (compare with the ionization parameter $\xi$,
\citealt{tatusaly}) $\tau_\mathrm{X}=\int_{R_*}^D\kappa\rho \mathrm{d}r \sim
\dot M(D-R_*)/(v_\infty DR_*)$ and corresponding parameters of Vela X-1 as
\begin{equation}
L_\mathrm{X}^\mathrm{max}e^{-\tau_\mathrm{X}}=
L_\mathrm{X}(\mbox{Vela X-1})e^{-\tau_\mathrm{X}(\mbox{\scriptsize Vela
X-1})},
\label{hloubkane}
\end{equation}
or, in scaled quantities
\begin{multline}
\log(L_\mathrm{X}^\mathrm{max}/\mbox{1 erg})=32.6+3.9
{\left({\frac{\dot
M}{1.5\times10^{-6}\,\mathrm{M}_{\odot}\,\mathrm{year}^{-1}}}\right)}\\
\times
{\left({\frac{v_\infty}{750\,\mathrm{km}\,\mathrm{s}^{-1}}}\right)^{-1}
\left({\frac{(DR_*)/(D-R_*)}{68.5\,\mathrm{R}_\odot}}\right)}^{-1}.
\label{zakladnismernice}
\end{multline}
Here $\dot M$, $v_\infty$,
$R_*$,
and $D$ are the wind mass-loss rate, the terminal
velocity,
radius of luminous component,
and the binary separation for individual HMXBs.

The results of observations for other HMXBs collected in Fig.~\ref{hmxb} clearly
support the picture that there exists a maximum allowed X-ray luminosity, which
depends on the wind and geometry parameters. Some systems lie close to the
boundary of the forbidden area, whereas others have typically higher wind
mass-loss rate $\dot M\gtrsim 5\times10^{-6} \,\text{M}_\odot
\,\text{year}^{-1}$ that keeps them out of the boundary. The position of
individual stars in this diagram may vary with time due to the existence of two
possible solutions of wind equations (as discussed in Sect.~\ref{nasebista}).

\subsection{Mass-loss rate determination}

Our results support the mass-loss rate predictions based on up-to-date wind
models in two ways. First, our mass-loss rate prediction of HD\,77581,
$1.5\times10^{-6}\,\mathrm{M}_{\odot}\,\mathrm{year}^{-1}$, agrees with the
value estimated from X-ray spectroscopy,
$1.5-2\times10^{-6}\,\mathrm{M}_{\odot}\,\mathrm{year}^{-1}$ \citep{viteal}.
Moreover the wind mass-loss rate cannot be lower than this value, because a
decrease of the wind mass-loss rate would lead to lower X-ray opacity, stronger
wind X-ray photoionization close to the star, and even more significant
reduction of the radiative force. This would finally cause a disruption of the
wind and a disappearance of X-ray emission. This imposes a strong lower limit on
the observational wind mass-loss rate estimates, that is in agreement with
current mass-loss rate predictions \citep{vikolamet,cmf}.

\subsection{Limitations of the present models}

The consistent inclusion of the X-ray irradiation, which is an advantage of our
models, also determines their shortcomings. The coupled solution of NLTE and
radiative transfer equations is significantly time-consuming even in 1D. Its
inclusion into multidimensional time-dependent simulations is likely beyond the
possibilities of current computers. Consequently, while one part of the problem
solution is treated in detail, the second one is simplified. The correct picture
of the flow in the HMXBs should take into an account the results of both
approaches: the stagnation of the flow in the direction towards the neutron
star, which is studied in this paper, and a complex 2D picture of the wind
accretion on the neutron star \citep{blok,blowoo,felan}.

On the other hand, there are effects that are not described by any of the
available models. It is well established that the hot star wind is inhomogeneous
on small scales (clumped, see \citealt{potclump}). In the case of HMXBs,
clumping (which favors recombination) may affect the region in which the
photoionized bubble is formed \citep{lidazasetaky}. However, the hydrodynamical
simulations \citep{felpulpal,runow} predict that clumping starts above the
critical point of the wind solution, consequently, it does not affect wind
mass-loss rates and terminal velocities. Thus we expect that clumping does not
significantly influence the results of our models.

The wind inhomogeneities likely cause the high variability of the X-ray source
\citep{prvni,lidazasetaky}. The calculation of the wind ionization should in
fact account for such time-dependent X-ray photoionization. However, because the
typical flow time of the wind is longer than the typical timescale of X-ray
variability, we expect that our models are able to reproduce the mean effect of
the X-ray photoionization. On the other hand, the variable X-ray ionization
source causes temporal changes of the radiative acceleration, providing external
perturbation which may contribute to the natural wind clumping. As a result, the
primary wind may be more clumpy than the wind of a similar single supergiant.

\section{Conclusions}

We provide detailed numerical models of the influence of X-rays on the
supergiant wind in the Vela X-1 binary system. The effect of X-ray
photoionization on the radiative force and wind dynamics has never been studied
using appropriate NLTE wind models. The X-rays photoionize the wind and destroy
the ions responsible for the wind acceleration. This results in flow stagnation
in the vicinity of the neutron star, which was identified in observations.
%
For a sufficiently strong X-ray source the wind that directly faces
the neutron star falls back on the mass losing star and never reaches the
compact companion.

We have shown that there is an upper limit to the X-ray luminosity the compact
star can have without disrupting the stellar wind. For a higher luminosity than
the limiting one the decrease of the wind acceleration is so strong that no wind
material would reach the neutron star. This theoretical picture of the maximum
X-ray luminosity is supported by observation of many high-mass X-ray binaries.

The wind equations allow the existence of two types of solutions with different
X-ray luminosities and wind velocities. The case of a strong X-ray source, which
significantly affects the wind ionization, leads to accretion of slow wind in
large amounts resulting in a strong X-ray source. On the other hand, a weak
X-ray source that does not significantly influence the wind ionization results
in accretion of fast wind in low amounts producing a weak X-ray source.
Different types of solution may appear in different binary systems, or
perturbations may cause switching between these two types of wind solutions
contributing to the X-ray variability.

The predicted mass-loss rate agrees with the value estimated from X-ray
spectroscopy. Moreover, the wind mass-loss rate cannot be lower than this value,
because a decrease of the wind mass-loss rate would lead to the disruption of
the wind and disappearance of the X-ray emission. This supports the reliability
of current mass-loss rate predictions.

\acknowledgments
We thank an anonymous referee for valuable comments on the manuscript. This work
was supported by grants GA \v{C}R 205/08/0003 and MUNI/A/0968/2009.

\end{document}